\pdfoutput=1
\documentclass[11pt]{article}

\usepackage{graphicx}
\usepackage{braket}

\usepackage{amsmath,amssymb,amsbsy,amstext, amsthm, simplewick}

\usepackage{cite}
\usepackage{amsfonts}
\usepackage{amssymb}
\usepackage{mathtools}
\usepackage{colortbl}
\definecolor{lightgreen}{cmyk}{0.2, 0, 0.2, 0.2}
\definecolor{lightgray}{cmyk}{0.1,0.2,0,0.1}
\definecolor{lightgray2}{cmyk}{0.1,0.1,0,0.1}

\makeatletter
\newlength{\apb@width}
\newcommand{\autoparbox}[2][c]{\settowidth{\apb@width}{#2}\parbox[#1]{\apb@width}{#2}}

\makeatother

\numberwithin{equation}{section}

\def\beq{\begin{equation}}
\def\eeq{\end{equation}}

\def\bea{\begin{eqnarray}}
\def\eea{\end{eqnarray}}

\def\beq{\begin{equation}}
\def\eeq{\end{equation}}
\def\be{\begin{equation}}
\def\ee{\end{equation}}
\def\bea{\begin{eqnarray}}
\def\eea{\end{eqnarray}}

\def\0{{\vec{0}}}

\DeclareRobustCommand{\SkipTocEntry}[4]{}

\def\beq{\begin{equation}}
\def\eeq{\end{equation}}

\def\ba#1\ea{\begin{align}#1\end{align}}
\def\bg#1\eg{\begin{gather}#1\end{gather}}
\newcommand{\bseq}{\begin{subequations}}
\newcommand{\eseq}{\end{subequations}}

\renewcommand{\ln}{\mathop{\rm ln}\nolimits}

\DeclareSymbolFont{extraup}{U}{zavm}{m}{n}
\DeclareMathSymbol{\varheart}{\mathalpha}{extraup}{86}
\DeclareMathSymbol{\vardiamond}{\mathalpha}{extraup}{87}

\def\({\left(}
\def\){\right)}
\def\[{\left[}
\def\]{\right]}

\begin{document}

\begin{titlepage}

\setcounter{page}{1} \baselineskip=15.5pt \thispagestyle{empty}

\vbox{\baselineskip14pt
%\hbox{hep-th/0000000}
}
{~~~~~~~~~~~~~~~~~~~~~~~~~~~~~~~~~~~~
~~~~~~~~~~~~~~~~~~~~~~~~~~~~~~~~~~
~~~~~~~~~~~ }

\bigskip\
%\hbox{CALT-TH-2019--031}
\vspace{1cm}
\begin{center}
{\fontsize{17}{36}\selectfont  
{
Detuning the BSW Effect with Longitudinal String Spreading
%BSW detuning
}
%SITP-18/02
}
\end{center}

\vspace{0.6cm}

\begin{center}
Dayshon Mathis
\end{center}

%\vspace{0.2cm}

\begin{center}
\vskip 8pt

\textsl{
\emph{Stanford Institute for Theoretical Physics, Stanford University, Stanford, CA 94306, USA}}
%\vskip 7pt
%\textsl{\emph{$^2$Walter Burke Institute for Theoretical Physics, California Institute of Technology, Pasadena, CA 91125, USA}}
%\vskip 7pt
%\textsl{\emph{$^3$Institute for Quantum Information and Matter, California Institute of Technology, Pasadena, CA 91125, USA}}
%\vskip 7pt
%\textsl{ \emph{$^4$Centro At\'omico Bariloche and CONICET, Bariloche, Argentina}}

\end{center}

\vspace{0.5cm}
\hrule \vspace{0.1cm}
\vspace{0.2cm}
{ \noindent \textbf{Abstract}
\vspace{0.3cm}

Black holes are interesting astrophysical objects that have been studied as systems sensitive to quantum gravitational data. The accelerated geometry in the exterior of extremal black holes can induce large center-of-mass energies between particles with particular momenta at the horizon. This is known as the Bañados-Silk-West (BSW) effect. For point particles, the BSW effect requires tuning to have the collision coincide with the horizon. However, this tuning is relaxed for string-theoretic objects. String scattering amplitudes are large in the Regge limit, occurring at large center-of-mass energies and shallow scattering angles, parametrically surpassing quantum field theoretic amplitudes. In this limit, longitudinal string spreading is induced between strings with a large difference in light-cone momenta, and this spread can be used to ‘detune’ the BSW effect. With this in mind, quantum gravitational data, as described by string theory, may play an important role in near horizon dynamics of extremal Kerr black holes. Further, though it may be hard to realize astrophysically, this system acts as a natural particle accelerator for probing the nature of small-scale physics at Planckian energies.

\vspace{0.4cm}

\hrule

\vspace{0.6cm}}
\end{titlepage}

\section{Introduction}

The Kerr metric, a description of the exterior space-time of a spinning black hole, was discovered in 1963, and the Penrose effect was deduced in 1969. The Penrose process is the phenomenon in which a particle decays into debris with particular kinematics in the ergosphere of a spinning black hole, leading to an ejected decay product with a large amount of energy. However, there is a $2\rightarrow 2$ collisional version of the Penrose process \cite{piran}, which produces a large center of mass energies analogous to particle accelerators. Extremal Kerr black holes can generate arbitrarily large center of mass energies \cite{Jacobson} \cite{Banados2009} if one of the infaller's trajectories is tuned to be \textit{critical} \cite{Zaslavskii2011}\cite{Zaslavskii2011kin}\cite{zaslavskiiPhysRevD.82.083004}, this is known as the Bañados-Silk-West (BSW) effect. A critical trajectory has an energy-angular momentum ratio that matches the angular speed of the space-time at the horizon, $E/L = \omega_H$. In the extremal limit $a\rightarrow M$, critical trajectories with angular momentum, $L$, have energy equal to the peak of the effective potential, $V_{\text{eff}}(L)=E$. The peak of the effective potential for relativistic orbits coincides with the horizon in the extremal limit, and critical infallers reaching the horizon have vanishing Boyer-Lindquist radial momentum. However, their local proper velocities are less than the speed of light. This is contrary to usual trajectories with finite radial momentum and local velocity equal to the speed of light at the horizon. With redshifting factors, the momenta of usual trajectories diverge at the horizon. In contrast, critical momenta are finite, leading to the collision between a critical and a usual particle having arbitrarily large center-of-mass energy. 

Assuming the infallers are colliding point particles implies the space-time coordinates must coincide upon collision. This requires tuning the timing of at least one of the trajectories, not only tuning the energy-angular momentum ratio. However, objects in string theory have a large extent at high energies (i.e., large $p^+$ in the light-cone coordinates) and spread longitudinally \cite{PhysRevD.49.6606} \cite{Dodelson:2015toa} \cite{dodelson2015} and with higher amplitudes than QFT models 
\begin{align}
    \langle \Delta X^+ \rangle~ \sim \frac{p^+_D}{p_{\perp}^2 + m_D^2 + \frac{1}{\alpha'}}
\end{align}
if we are in the Regge limit $s \rightarrow \infty$ and $t \rightarrow \text{fixed}$. In \cite{Dodelson2017} \cite{Mousatov:2020ics}, these large momenta strings are seen as detectors sensitive to early infaller data. Further, the spreading is not a gauge artifact. By analysis of the six-point amplitudes \cite{Dodelson:2017hyu}, the spreading still manifests. Analogously, usual particles have large $p^+$ and, consequently,  large longitudinal spread, making usual particles sensitive to critical infallers and relaxing the need for tuning the timing. We could assume a statistically large amount of particles are accreting into the hole to get detection and detuning for point particles, but critical trajectories are still tuned. In the extremal case,  the innermost stable circular orbit (ISCO) coincides with the horizon, possessing a critical energy-angular momentum ratio. ISCOs have the largest binding energy, so infallers undergoing radiative processes naturally reach ISCO with critical momenta.  Strings, like point particles, are driven by the Kerr geometry to criticality. Strings zero-mode undergo similar point particle phenomena; additionally, objects of finite extent around Kerr holes go through a stringy Penrose process, where strings gain or lose energy by stretch as opposed to releasing ejecta \cite{Igata2016}\cite{Igata2018}. %Critical trajectories don't seem to change the extremality of the EK hole and seem like a representation of long-lived states or EK BH microstates.

Since the extremal Kerr exterior geometry induces large COM energy, it acts as an effective particle accelerator to probe the quantum nature of gravity. However, extremal Kerr black holes may be non-astrophysical due to the Thorne limit \cite{thorne1974}, back-reactions, and other limitations \cite{Berti2009}. There is a chance to overcome these limits \cite{Sdowski2011}\cite{Tanatarov2013}, but regardless of the physical hindrances an extremal Kerr background will be assumed for its theoretical interest.

%Section summary
Section 2 reviews the BSW effect and determines the near-horizon kinematics. Next, in the second section, the string theoretic version is analyzed for the sake of detuning the BSW effect with the non-local spread of the string form factor and a Penrose process for an extended object. The third section qualifies some of the limitations and concludes the paper. Further, if outgoing strings from the collisions can escape the hole, this could leave a quantum gravity signature on quasinormal mode.

\section{Kerr Kinematics}  \label{sec:kinematics}
The Kerr metric in Boyer-Lindquist coordinates ($G = c = 1$)\cite{boyer1967} is 
\begin{align}\label{eq: metric}
    ds^2 = -N^2 dt^2 + g_{\phi \phi}(d\phi - \omega dt)^2 + dl^2 + g_{\theta\theta} d\theta^2,
\end{align}
where $N$ is the lapse function 
$$N^2= \frac{\Sigma \Delta}{(r^2+a^2)^2 - a^2 \Delta \sin^2 \theta}~,$$
$\omega$ is the angular speed on a co-rotating frame at a Boyer-Lindquist radius, $r$, 
$$\omega = \frac{2 a M r}{(r^2+a^2)^2-a^2 \Delta \sin^2 \theta},$$
an d$l$ is the proper distance $l = \int dr \sqrt{\frac{\Sigma}{\Delta}}$, $\Sigma = r^2 + a^2\cos^2 \theta$, and $\Delta = r^2 + a^2-2Mra$.  A light-cone coordinate, the $x^\pm$-direction is be defined
\begin{equation}
    dx^\pm = \frac{Ndt\pm dl}{\sqrt{2}}.
\end{equation}

When considering infallers, for simplicity, we will consider equatorial motion, $\theta = \frac{\pi}{2}$. However, incoming particles are naturally driven to equatorial motion by the Bardeen-Peterson effect\cite{bardeenpeterson} (although non-equatorial motion exhibits the BSW effect\cite{Zaslavskii2012}). The equation of motion for a particle with mass, $\mu$, is
\begin{align}
    \sqrt{2} p^\pm /\mu &= \frac{E-\omega L}{N} \pm \sqrt{(\frac{E-\omega L}{N})^2-(\delta +\frac{L^2}{g_{\phi\phi}})}\\
    (p^\phi - \omega p^t)/\mu &= \frac{L}{g_{\phi \phi}}\\
    p^\theta = 0
\end{align}
where $E$ and $L$ are the energy per mass and angular momentum per mass, respectively, and the value of $\delta$ depends on the geodesic of the particle being space-like ($\delta=-1$) or time-like ($\delta=1$). The light-like case ($\delta=0$) requires the momenta per mass to become coordinate-derivatives of an affine parameter $p^\mu \rightarrow \nu \frac{dx^\mu}{d\lambda}$ and $E$ and $L$ represent energy per frequency and orbital angular momentum per frequency \cite{zaslavskiimassless}. As argued in \cite{zaslavskii2010} \cite{Tanatarov2012} \cite{MedvedVisser2004}\cite{Medved2004}, the angular frequency in the near horizon extremal limit
\begin{equation}
    \omega = \frac{1}{2M} - \frac{N}{M} + \mathcal{O}(N^2),
\end{equation}
the non-extremal angular speed deviation from the horizon speed is quadratic in the lapse function. Seen, the expansion of $\omega$ for the non-extremal case in terms of $r-r_+$ is the same as the extremal case, however in the extremal case $N^2 \sim (r - r_+)^2$ while the non-extremal case has $N^2 \sim (r-r_+)(r_+-r_-)$. This expansion means that the momentum at the horizon is finite for a critical particle ($E - \omega_H L =0$) around an extremal hole.
\begin{align}
     \sqrt{2} p^+_C /m_C  = 2\frac{L_C}{M}\\
    \sqrt{2} p^-_C /m_C  = \frac{M(\delta_C +\frac{L^2_C}{g_{\phi\phi}})}{2L_C}
\end{align}
whereas for horizon momenta of usual particles ($E_U-\omega L_U\neq 0$), $p^+$ diverges
\begin{align}
    \sqrt{2} p^+_U /m_U  = 2\frac{E_U-\omega_H L_U}{N}\\
    \sqrt{2} p^-_U /m_U  = \frac{N(\delta_U +\frac{L^2_U}{g_{\phi\phi}})}{2(E_U-\omega_H L_U)}
\end{align}
Calculating the center of mass energy, $s = -(p_U + p_C)^2 \sim -p^+_Up^-_C$, leads to divergent COM energy, $s \sim \mathcal{O}(\frac{1}{N})$. This is the essence of the BSW effect. The large COM energy theoretical can reach Planckian energies, lending credence to the necessity of quantum gravity data for near horizon extremal black holes physics.

There is a large amount of proper distance at the horizon, although it is not evident in Boyer-Lindquist coordinates. This leaves plenty of space to explore string spreading in the near horizon region. Calculating the proper distance between relativistic orbits shows the large proper length in the near horizon region. Infalling critical particles asymptotically rests on the unstable turning point in the extremal limit. These turning points for positive energy critical particles exist between the innermost-stable-circular orbit (ISCO)
\begin{align}
    r_{ISCO} &= M\{3+Z_2 - \sqrt{(3-Z_1)(3+Z_1+2Z_2)}\}\\
    Z_1 &= 1+(1-\frac{a}{M}^2)^{1/3}\{(1+\frac{a}{M})^{1/3}+(1-\frac{a}{M}\}{M})^{1/3})\\
    Z_2 &= \sqrt{3\frac{a}{M}^2+Z_1^2}
\end{align}
and the circular photon orbit (CPO)
\begin{align}
    r_{CPO} = 2M\{1+\cos (\frac{2}{3}\cos^{-1}(-\frac{a}{M}))\}.
\end{align} 
The proper distances between ISCO, the marginally bound circular orbit ($E = \mu$) (MBCO),
\begin{align}
    r_{MBCO} = M\{1+\sqrt{1-\frac{a}{M}}~\}^2
\end{align}
CPO and the proper distance from the horizon to all these radii are all non-zero. Nearing extremality, $a = 1 - \epsilon$, the behaviour of these radii go as
\begin{align}
    r_+ = M + (2\epsilon)^{1/2}\\
    r_{\text{CPO}} = M + (\frac{8}{3}\epsilon)^{1/2}\\
    r_{\text{MBCO}} = M + 2\epsilon^{1/2}\\
    r_{\text{ISCO}} = M + (4\epsilon)^{1/3}
\end{align}
and with the proper distance at the horizon being
\begin{equation}
    l_{1, 2} \sim r_+ \log{\frac{r_2-M}{r_1-M}}
\end{equation}
which is large as extremality is approached. Assuming the extremal Kerr geometry maintains $a = M$, there is a large amount of space at the horizon, which leaves string extent analysis unhindered by the background.

It can also be shown the energy-angular momentum ratio required to rest at this turning point is equivalent to criticality. A turning point placed at radius, $r$, between ISCO and CPO has energy and momentum
\begin{align}
    E =& \frac{1-2M/r+aM^{1/2}/r^{3/2}}{\sqrt{1-3M/r+2aM^{1/2}/r^{3/2}}}\\
    L =& \frac{M^{1/2}r^{1/2}-2aM/r+M^{1/2}a^2/r^{3/2}}{\sqrt{1-3M/r+2aM^{1/2}/r^{3/2}}}.
\end{align}
and in the near horizon extremal limit $\frac{E}{L}\rightarrow \omega_H$. Critical particles can be considered to be stuck near, have fallen from, or are approaching its centrifugal barrier, so the conserved energy and angular momentum, $E$ and $L$, are appropriate for understanding infallers near the horizon. A large amount of regular space at the horizon leaves plenty of room for scattering dynamics and string theoretic phenomena. The space near the horizon is flat and tidal forces are weak for large mass black holes, $M >> m_P, m_{\text{string}}$.

The gist of the BSW effect is the generation of a diverge, $\mathcal{O}(N^{-1})$, COM energy, induced by the geometry, between colliding usual and critical trajectories. Let the momentum of the usual particle near an extremal horizon be denoted by $p_U$ and the momentum of the critical particle $p_C$.  The center of mass energy is therefore
\begin{align*}
    s &= -(p_U+p_C)^2\\
    &= -p^+_Up^-_C - p^+_Cp^-_U + g_{\phi \phi}(p^\phi_C - \omega p^t_C)(p^\phi_U - \omega p^t_U)  + m^2_C + m^2_U.
\end{align*}
 Since $p^+_U$ is large near the horizon, the dominant term is 
\begin{align}
     s \sim m_U m_C \frac{M(E_U-\omega_H L_U)(\delta_C + \frac{L^2_C}{g_{\phi \phi}})}{N L_C} + \mathcal{O}(N^0).
\end{align}
Now, with a COM energy that can access Planckian energies, we can analyze string amplitudes in this regime. The Regge limit, $s \rightarrow \infty$ with $t \rightarrow \text{fixed}$, has a large scattering amplitude for interacting strings. The Regge limit corresponds to very small scattering angles, but the Boyer-Lindquist coordinates obscures angular separation, so 

For $2\rightarrow2$ scattering, the contraction of momenta in the final state must also be large by conservation of momentum, so a usual-critical pair in the final state is a good assumption.
In the case where we have usual-critical $\rightarrow$ usual-critical,  the Mandelstam variable $t = -(p_C - p_{\hat{C}})^2$ is
\begin{align}\label{eq:tCC}
    t = m_{\hat{C}} m_C(\frac{\delta_C L_{\hat{C}}}{L_C} +  \frac{\delta_{\hat{C}}L_{C}}{L_{\hat{C}}}) + m_C^2 + m_{\hat{C}}^2,
\end{align}
and $u = -(p_C - p_{\hat{U}})^2$ can be solved with $s+t+u= \sum_i m_i^2$ and is also large due to the contraction between usual and critical momenta. The final state constants, $E_{\hat{C}}$, $E_{\hat{U}}$, $L_{\hat{C}}$, and $L_{\hat{U}}$ are solved by $p_C^\mu + p_U^\mu = p_{\hat{C}}^\mu + p_{\hat{U}}^\mu$ and we'll assume for simplicity the final state particles are also equatorial. To be sure we are in the stringy Regge regime we impose 
\begin{align} \label{eq: kinematics cond}
    \text{sgn}(p^l_C) &= \text{sgn}(p^l_{\hat{C}}) = \sigma_C \\
    p^l_{U} &< 0 \nonumber\\
    p^l_{\hat{U}} &< 0 \nonumber.
\end{align}
This leads to the final state constants being
\begin{align}
    &E_{\hat{U}} = E_U +  3 \omega_H L_C + 2\omega_H\sigma_C\sqrt{3L_C^2-4\delta_CM^2}  ...\nonumber\\
    &... \pm\sqrt{21L_C^2+12 \sigma _C L_C  \sqrt{3L_C^2-4\delta_C M^2} -16 M^2 \delta_C}+ \mathcal{O}(N)\\
     & E_U + E_C = E_{\hat{U}}+E_{\hat{C}}\\
     & L_U + L_C = L_{\hat{U}}+L_{\hat{C}}
\end{align}
and particle-$\hat{C}$ is critical, $E_{\hat{C}} - \omega_H L_{\hat{C}} = 0$. In this case, we are looking at the Regge regime since $ t =\text{fixed}$ as seen from \eqref{eq:tCC}. In the Regge limit, string amplitude is large compared to the QFT amplitude. Additionally, other stringy objects have an extent (open/closed string, D-brane, superstrings, etc.) and exhibit the same spread longitudinal spread in the Regge limit. Hence, this analysis is broadly applicable to string theory.

%Branes as a description for extremal black holes (add note about superstring haveing regge limits)
It'd be interesting to study brane dynamics deep in the throat and elsewhere in the context of the BSW effect since brane-brane and string-brane amplitudes also exhibit the Regge behavior. Branes have been used in realizing the microscopic of extremal black holes in higher dimensions \cite{Strominger1996}, so the BSW effect inducing longitudinal string spreading could become relevant with the growth of an extremal black hole from a usual infaller or interactions with past infallers.

\section{Detuning the BSW effect with Longitudinal String Spreading}
The amplitudes for interacting strings contain the convolution of vertex operators, $V(z, \bar{z})$. An open/closed string tachyon has a vertex operator of the form ($\hbar = 1$)
\begin{align}
    V \sim 
    \frac{g_{o, c}}{\alpha'} \int d^2 z~ e^{i p \cdot X(z, \bar{z})}.
\end{align}

 Higher excited states terms have a more involved vector/tensor mode construction, but the required Regge behavior manifestation can be deduced from the simplified tachyon vertex operator. The four-point tree-level amplitude is a convolution of these vertex operators
 \begin{align}\label{eq:vertexamp}
     A = \frac{g^{-\chi}}{\text{Vol}} \int d^2 z_U d^2 z_C d^2 z_{\hat{C}} d^2 z_{\hat{U}} \langle V_C(z_C) V_U(z_U) V_{\hat{C}}(z_{\hat{C}}) V_{\hat{U}}(z_{\hat{U}}) \rangle.
 \end{align}
 The $\text{SL}(2, \mathbb{C})$ group reduces the integrand so that we consider, $V_U(z_U)$ and 
$V_C(0)$. Regardless of the type of stringy objects involved, the integrand possesses a term
\begin{align}
    e^{ip_U\cdot X(z_U, \bar{z}_U)}e^{ip_C \cdot X(0)},
\end{align}
which elicits the Regge behavior when evaluated through the operator product expansion,  
\begin{align}
     |z_U|^{-\frac{\alpha' t}{2}-4}e^{i(p_U+p_C)\cdot X(0)+p_U \cdot (z_U \partial + \bar{z}_U\bar{\partial})X(0)}.
\end{align}
Integrating over the world sheet leads to the Pomeron vertex amplitude as seen in \cite{Brower2007}
\begin{align}\label{eq:pomeronintegrand}
    \sim 2\pi e^{-i \pi(\alpha' t/4 + 1)}
    \frac{\Gamma(-1-\alpha't/4)}{\Gamma(2+\alpha't/4)} e^{i(p_U+p_C)\cdot X(0)}(p_U\cdot \partial X(0) ~p_U\cdot \bar{\partial}X(0))^{1+\alpha' t/4}.
\end{align}
where $\Pi(\alpha' t) = 2\pi e^{-i \pi(\alpha' t/4 + 1)} \frac{\Gamma(-1-\alpha't/4)}{\Gamma(2+\alpha't/4)}$, and putting equation \eqref{eq:pomeronintegrand} back into equation \eqref{eq:vertexamp} give the $s$-dependence which exhibits the large form factor
\begin{align}
\mathcal{A} \sim \Pi(\alpha' t)(e^{-i\pi /2} \alpha' s /4) ^{2+\alpha' t /2}
\end{align}

In the Regge limit $s\rightarrow \infty$ $t \rightarrow \text{fixed}$, the result is
\begin{align}
    \mathcal{A}\sim (\frac{\alpha' s}{4})^2 \frac{\Gamma(-1-\alpha' t /4)}{\Gamma(2+\alpha' t /4)} (e^{-i\pi /2} \alpha' s /4) ^{\alpha' t /2}
\end{align}
which has the modified form factor, $s^{\alpha' t /2}$. Replacing the Mandelstam variable, $t$, with the transverse momentum-squared, $q_\perp^2 = -t$, shows the form factor becomes
\begin{align}
    e^{-\frac{\alpha' q^2_\perp}{2}\ln(\alpha' s)}
\end{align}
with the logarithmic piece in the exponent contributing to the effective growth of the impact parameter. More complicated string interactions also have a similar amplitude in the Regge limit, exhibiting the logarithmic transverse spread and, as seen below, the large longitudinal spread.

%comparing string hard scattering, string regge limit, and QFT amplitude
With the Kerr near-horizon region being flat compared to the string scale, we can consider large transverse separation $\Delta X_\perp ^2 \sim \alpha' \log (s)$ where QFT amplitudes are suppressed, but string amplitudes are relatively large. The transverse spread $$\langle (\Delta X _\perp)^2 \rangle \sim \alpha' \log (s/t) \sim \mathcal{O}(\log(N))$$ is also large compared to QFT fields interacting on tails of their wave function. The hard scattering counterpart with finite angle difference between $U$ and $\hat{U}$, $s\rightarrow \infty$ and $\frac{t}{s} = \text{fixed}$, has an amplitude that goes like $e^{-\alpha's}$ which has a softer falloff-- suggesting an object of size $\sqrt{\alpha'}$, while the Regge amplitude is large in $x_\perp^2 \sim \alpha' \log (s)$. The large amplitude is one reason to use the Regge limit to probe information in a quantum gravitation context, along with the gravitational information in the $t \rightarrow 0$ limit.

The BSW effect implies string physics in a certain kinematic regime, at least, with fixed Mandelstam variable, $t\rightarrow \text{fixed}$. The longitudinal string spreading is the primary effect that relaxes the need for tuning the initial conditions to guarantee a near-horizon collision. We now turn to the longitudinal string spreading effect that detunes the timing necessity. As seen in \cite{Dodelson:2015toa} \cite{Dodelson:2017hyu}, 
\begin{align}\label{eq:longspread}
    \langle \Delta X^+ \rangle~ \sim \frac{p^+_U}{p_{U\perp}^2 + m_U^2 + \frac{1}{\alpha'}}
\end{align}
This comes directly from the expectation value of the $X^+ - x_{cm}^+$ for an open string in the light-cone gauge, $X^- = x^- + 2\alpha^-\tau$,
\begin{align}
    X^+(\tau=0) = x^+_{cm} + i\sqrt{2\alpha}\sum_{n \neq 0}\frac{1}{n}\alpha^+_n \cos(n\sigma)\\
    X^i(\tau=0) = x^i_{cm} + i\sqrt{2\alpha}\sum_{n \neq 0}\frac{1}{n}\alpha^i_n \cos(n\sigma)
\end{align}
calculating the variance for both, $\braket{(X^\mu-x^\mu)^2} $ and implementing the Virasora algebra, 
\begin{align}
    [\alpha^+_n, \alpha_m^+] = \sqrt{\frac{2}{\alpha}}\frac{(m-n)}{p^-}\alpha^+_{m+n} + \frac{4m^3}{\alpha'(p^-)^2}\\
    [\alpha^i_m, \alpha^j_n] = m \delta_{m+n,0}\delta^{ij}
\end{align}
leading to the variances,
\begin{align}
    \braket{(\Delta X^+)^2} \sim \frac{1}{(p^-_C)^2}\sum_{n>0}^{n_{\text{max}}}n\cos^2n\sigma\\
    \braket{(\Delta X^i)^2} \sim \alpha' \sum_{n>0}^{n_{\text{max}}}\frac{1}{n}\cos^2n\sigma.
\end{align}
The maximum mode, $n_{\text{max}}$, to which the detector, $U$, is sensitive, in the Regge limit,  is given by \cite{Dodelson2017}
\begin{align}
    n_{\text{max}}\sim \frac{p^-_Cp^+_U}{p_\perp^2 + m^2_U + \frac{1}{\alpha'}}.
\end{align}
The result is equation \eqref{eq:longspread} and a logarithmic dependence for the transverse mode
\begin{align}
    \braket{\Delta X _\perp^2} \sim \alpha' \log(\frac{s}{t}).
\end{align}
Longitudinal spreading is proportional to the light-cone momenta of string $U$. With large $p^+_U$, longitudinal spreading is also very large, spreading in the radial direction, relaxing the tuning conditions.

%time resolution from small worldsheet parameter
 Although the string spreading is large, the interaction involves small regions on the worldsheet. The interaction on the world sheet has a large contribution around a point, $z\sim \frac{t}{t+s}$. This saddle point is small in the Regge limit, since $s\rightarrow \infty$ and $t\rightarrow \text{fixed}$. This is related to the small time resolution of string $U$; since $p^+_U$ is large, the internal clock of $U$ can resolve string excitations on string $C$.

\section{Discussion of Limitations}

\subsection{COM energy, timing, and redshift}
The divergent COM energy generated by the Kerr geometry can be bounded in a string theory \cite{DING2013}. Since string theory generates a non-commutative geometry, the Kerr black hole develops a mass distribution. The mass is an incomplete gamma function
\begin{align}
    M(r) = \frac{2 M_0}{\sqrt{\pi}}\int^{\lambda \frac{r^2}{\alpha'}}_0dt t^{1/2}e^{-t}
\end{align}
computing the BSW effect with this addition to the metric bounds the COM energy to of the order of the black hole mass, $\sqrt{s} \leq \mathcal{O}(M_0)$. This new bound is reasonable and implies the Regge limit intermediate string can theoretically have a large portion of the black hole's mass. This is promising for the strings involved in a Blanford-Znajek-like process.

The presence of the BSW effect leads to implications on complementary and UV-completion. Understanding where the BSW effect can arise and the limitations suppressing large COM energies is essential.

Many limitations in the literature haven't been considered for stringy processes. The detuning of parameters inducing the BSW effect also relaxes some suppressive effects. The critical particle may take a long time to reach the horizon (disallowing multiple scattering) and fall on a floating orbit where the radiation to infinity matches and changes the extremality of the hole along the way. With longitudinal spreading, the ample proper time to reach BSW can also be relaxed by string spreading.

It may take a large amount of time for critical particles from infinity to reach the collision zone. It is not apparent how string spreading can resolve this since spreading occurs in the near-horizon region, far from $r = r^+ + \delta$. The breakdown of EFT here requires a long waiting time, like the Schwarzschild rapidity-though-time-displacement calculation.

Events near the horizon are red-shifted before escaping, which may spoil the chance for a direct string signature in the overtones of QNM.  The COM energy is controlled by the near extremal spin parameter $a = 1- \epsilon$. For a BSW event occurring on $r_+$, ISCO, MBCO, and CPO have the following dependence on $\epsilon$ \cite{pradhan2016extremal},
\begin{align}
    s_{\text{ISCO}} \sim \frac{1}{\epsilon^{1/3}}\\
    s_{\text{MBCO}} \sim \frac{1}{\epsilon^{1/2}}\\
    s_{\text{CPO}} \sim \frac{1}{\epsilon^{1/2}}\\
    s_{r^+} \sim \frac{1}{\epsilon^{1/2}}.
\end{align}

In \cite{Schnittman}, it is shown there are multiple scattering events that can enhance the energy escaping the near horizon geometry at the cost of spinning down the black hole. However, the spin-down of an extremal hole to sub-Planckian BSW event takes over around the Hubble time for supermassive, $M\sim \mathcal{O}(10^8M_{\text{Sun}})$. The Hawking temperature is less than the CMB time, so the loss of extremality by Hawking radiation, especially for large holes, is negligible.

In \cite{Igata2018}, it is shown the induced metric from Nambu-Goto strings has similar (though not identical) evolution to magnetic `field sheet' around the rotating black hole. Strings like magnetic fields are slowed when $\frac{d\phi}{dt} >  \omega$ and accelerated by the geometry $\frac{d\phi}{dt} <  \omega$. This is another effect, facilitated by a Penrose-like process, that is useful for extended objects. The fine-tuning of critical particles is unnecessary for stringy particles, the geometry drives the string to have $\frac{d\phi}{dt}=\omega$, which matches the angular speed of the horizon. String zero modes with deviation from the co-rotating frame have non-zero energy flux scaled by the string tension. With a modest string coupling parameter, the geometry can induce string decay. Critical strings are not stretched nor drained by the geometry and fall unabated decay.

There are several future directions to take using the possible Planckian energy of a BSW event in the fiducial frame. This effect occurs for other extremal holes with exotic charges as well. Since the Regge amplitude is large for extended objects, branes-string or brane-brane scattering is large. In the Regge limit, gravitational data is probed, and an interesting direction may involve considering black holes source by string charges.

\section{Comments on the Implication of the BSW effect for extremal black holes}

%(In a sense, the nice slice argument for black hole non-locality is not sufficient to do the commutation of vector fields on nice-slices is non-vanishing but the non-local present in the gauge invariant S-matrix in the Regge limit supports non-locality at the horizon \cite{Lowe_1995}. The Pomeron vertex shows the longitudinal string spreading at the level of the S-matrix.)

\subsection{Signal out?: Comments on Quasinormal Modes and Superradiant modes}

Given the kinematics of the Regge limit described in \eqref{eq: kinematics cond}, where $\sigma_C=1$, particles do not easily escape the horizon regime. The final state critical particle has a chance to escape but would classically take a divergent amount of time to overcome its centrifugal barrier. The possibility of quantum tunneling past the barrier or multiple perturbing scatter processes could lead to the critical particle leaving a string signature in outgoing modes. The critical condition, $E_{\hat{C}} - \omega_H L_{\hat{C}}=0$, is reminiscent of the superradiant bound for quasinormal modes, $\nu - m \omega_H \leq 0$, so one might expect a stringy signature in SR mode \footnote{The is quasi-classical argument}. There is a large amount of work involving SR mode and the NHEK/CFT correspondence\cite{Guica2009} \cite{Bredberg2010}\cite{Gralla2016}. Analyzing the BSW effect and detuning from longitudinal string spreading in this context would be interesting to explore. 

Our Regge setup involves an infalling usual string, whose final state $\hat{U}$ is also plunging; however, outgoing usual strings can similarly exhibit Regge behavior. The usual trajectories may originate from deeper in the throat, $p^l_U >0$ and $p^l_{\hat{U}}$, which can possibly leave some signature in some outgoing quasinormal modes.

%Stringy BSW production may escape gravitational well and tunneling light particle
There is an additional issue, even if the data can escape the Kerr throat-- the redshift. The large gravitational redshift reduces the asymptotic local final state energy heavily. There is some tension regarding the strength of the redshift \cite{McWilliams2013} \cite{Bredberg2010} \cite{Zaslavskii2013} \cite{mcwilliams2013reply}. Regardless of the level of redshift, multiple scattering with particles or the barrier can induce Penrose processes (sometimes called the super-Penrose process for several collisions) that can increase the energy severalfold per interaction, and string data could escape with modest energy, assuming the quantum gravitation data survive the interactions.  The efficiency of each collision, $\eta_i = \frac{E_{\hat{C}, i}}{E_{D,i}+E_{C,i}}$, is large for extremal black holes and is divergent at CPO, $\eta_{\text{CPO}} \sim \frac{1}{\sqrt{1-a^2}}$ \cite{Schnittman}.  Near the horizon, the efficiency is $\eta = (2+\sqrt{3})^2 > 10$ and the COM energy diverges. The hole can donate energy and angular momentum to $\hat{C}$ but there is a finite number of scattering possible before the effect fails, and suggests this is 
\begin{align}
    N_{\text{scatter}} \sim \sqrt{\log_{10}(M/m_{\hat{C}})}.
\end{align}
For astrophysical black holes and initial electron-like masses, Planckian energies can extracted before the hole contracts \cite{Schnittman}. For point particles, tuning is required in these processes, but strings whose zero mode follows point particles (when tidal forces are weak) can reach these energies without such tuning. Multiple scattering requirements are related to an n-point scattering amplitude, $n > 4$, studied in \cite{Mousatov:2020ics}\cite{Dodelson2021}, and Regge behavior and longitudinal spreading are still present.

If $\hat{C}$ is a graviton, which is a non-coherent state,
\begin{equation}
    V \sim \frac{1}{\alpha'}\int d^2 z \sqrt{g} (h_{\mu \nu} \partial X^\mu \bar{\partial}X^\nu),
\end{equation}
while a coherent state is constructed from $e^V$. \cite{Parikh2021} shows gravitational wave interferometry can detect quantum gravity signatures. Stringy physics may be found in high-energy processes that produce QNM modes or late-time tails.

It is well-known that near the horizon of an extremal Kerr black hole is a CFT dual \cite{Guica2009}. The only scalar modes that survive are the superradiant ones with $\nu \hat{t} = m \hat{phi}$ (hatted variable are in NHEK coordinates) and $\nu \sim (E/m)_{\hat{C}}$. This is the SR bound. We expect the string's effective field decomposition to be
\begin{equation}
    \Psi = \int d\nu \sum_{l,m} R(\hat{r}) S_{s m l}(\theta, a\nu) \exp(i\nu \hat{t} - i m \hat{\phi})
\end{equation} 
but the NHEK limit seems to reject all modes except the superradiant ones. We therefore assume
\begin{equation}
     \Psi_{\hat{C}} =    R(\hat{r}) S_{s m l}(\theta, a\nu) \exp(-i\nu \hat{t} + i m \hat{\phi}) +\mathcal{O}(m_{\hat{C}}/E_{\hat{C}})
\end{equation}
where we ignore the effect of the mass. According to the EK geometry, this mode is exponentially excited from the hole and rejected, extracting energy. For $s = \pm 2$ and $a\nu \rightarrow \infty$, we have a peak at the south pole and north pole, respectively \cite{1978RSPSA.358..405C}. This may mean there is a stringy signature in jets. The local tidal force measurable at large $r$ by an interferometer is 
\begin{equation}
\mathcal{E}_{\hat{\theta}\hat{\theta}}   - i \mathcal{E}_{\hat{\phi}\hat{\theta}} \sim\frac{\sqrt{\nu^3 M}}{r}  S_{s m l}(\theta, a\nu) \exp(-i\nu \hat{t} + i m \hat{\phi} +i \nu r_{*})
\end{equation}
where the hatted coordinates are local. Deep horizon physics may be detected at large distances. The vertex operators of string theory are analogous to creation operators in QFT.

The Kerr background is known to be dual to a CFT representing information near the superradiant limit on near the extremal horizon. The implications of longitudinal spread translates to non-local distributions in the CFT. Quasinormal modes near the SR-limit corresponds to left-moving fields in CFT. It would be interesting to consider large longitudinal string spreading parameterise in terms of an outgoing signal.

The ignored mass, $m_{\hat{C}}$, may play an interesting role in the late-time tail of Kerr holes. The late-time function for massless perturbations is \cite{RevModPhys.83.793}
\begin{align}
    |\Psi| \sim t^{-(2l+3)}
\end{align}
whereas for massive perturbations,
\begin{align}
    |\Psi| \sim t^{-(l+3/2)}\sin(m_{\hat{C}, \hat{U}}t)
\end{align}
There may be high-frequency undulations in the late-time tail function.

\subsection{Large Strings}

The dynamics of large cosmic strings in a Kerr background have been of interest \cite{Frolov1996RigidlyRS} \cite{Xing2021}. These large stable strings, captured by a Kerr black hole, interact by sapping and imparting energy with the hole in a fashion similar to the Blanford-Znajeck process, but scales by the string tension, $\mathcal{O}(1/g\alpha')$ \cite{Igata2016}\cite{Igata2018}. These studies analyze the particular case of rigid strings, i.e., time-independent string configuration. Still, the intersection of the BSW effect, non-local longitudinal spreading, and large string may be an exciting avenue to explore in future works.

The regime of interest for perturbative string scattering scenarios is still at weak coupling, $g^2_o \sim g_c \ll 1$, as needed for the tree-level analysis above, and we can see from the Nambu-Goto action of an on-shell string the nature of the energy extraction/deposition. For stationary ($r = \text{cons}$) rigid strings with angular velocity $\Omega = \frac{E}{L}$, the induce metric with the static gauge $t = \tau$ is \cite{Igata2018}
\begin{align}
    h_{tt} = -N^2 + g_{\phi \phi}(\Omega -\omega)^2
\end{align}
 for critical infallers $\Omega = \omega_H$. When $h_{tt} = 0$ near the horizon, rigid data on the worldsheet hits a limit, where the time-translating killing vector is null-- the Killing horizon. The conserved momentum flux onto the world-sheet goes as $$\mathcal{P}_{\text{flux}} = \frac{1}{4\pi g \alpha'}\frac{d\sqrt{-\det h}}{d(d\phi/d\sigma)}$$ which is $\mathcal{O}(N)$ and exhibiting the tension scaling. A rigid critical string gains no energy from the black holes, while the usual rigid strings do the bulk of the energy extraction outside the throat where the lapse function is finite, $N^2 =  g_{\phi \phi}(\Omega -\omega)^2$. As seen in \cite{Igata2016}, rigid string around a hole is analogous to the Blanford-Znajek effect, which supplies an astrophysical jet. Therefore, an astrophysical jet may be another avenue for such quantum gravitational data. So,  energy is pumped into the string if it is too slow, $E/L < \omega_H$, and drained is $E/L>\omega_H$. Only critical strings are exempt from this energy exchange. Critical strings may naturally populate an equilibrium state, with the hole requiring even less tuning to attain critical momenta.

Although energy and momentum fluxed into the string are proportional to $N$, a $1/N$ amount of time needs to be spent in the throat to extract a significant amount of energy. Assuming stabilizing factors like weak coupling to suppress decay or warped tension, ala the Randall-Sundrum model, to get long-lived strings allowing for a stringy version of the Blanford-Znajek process. \iffalse The lifetime of a classical string is approximately $t_{\text{lifetime}} \sim \frac{1}{\Gamma_{\text{decay}}} = \frac{\alpha'_{\text{eff}}}{g^2 l_{\text{string}}}$. In contrast, a cosmic string's lifetime is inversely proportional to the tension $t_{\text{lifetime}} \sim \frac{\alpha'_{\text{eff}} l_{\text{string}}}{G_N}$. %from the Zwiebach text cosmic string analysis
 For the cosmic string, the estimate of the energy extraction is $\Delta E  \sim \frac{N l_{\text{string}}}{G_N}$. Interestingly, the energy extracted per unit length is independent of tension for cosmic string. With $l \sim E \alpha' = \mathcal{O}(\frac{1}{N})$\fi It seems the large string, in particular, has a chance to extract a large amount of energy from the hole. Further, considering the intermediate string, between a usual and critical string, as usual, it will also have divergent energy extracted from the hole. The Penrose effect is less efficient than its collisional counterpart, the BSW effect. However, the Penrose process may play a pivotal role in exciting decaying strings into black hole polar jets, analogous to the BZ process. As studied in \cite{Xing2021}, string theory signatures may uniquely be found in astrophysical jets and probe effective string tension at high energies.

%Large strings and Jets

\subsection{Tidal forces}

%recent work with divergent tidal force
In \cite{horowitz2023extremal}, it is shown adding curvature terms to the leading effective gravitational theory leads to diverging tidal forces consistent with analyticity, unitary, and causality, and supported by local string theory. The divergence in tidal force at the extremal Kerr horizon may be related to the divergence COM energy, a la the BSW effect. The context is unclear but it is a curiosity to see if there is more than a connection between divergence in tidal force from curvature terms and longitudinal spreading, as both have origins in a string theoretical context. This effect could happen in tandem with the BSW effect since modest acceleration does not spoil the BSW effect \cite{Tanatarov2013}.

It is natural for string theoretical black holes to be embedded in an asymptotic AdS space-time. Extremal BHs in this context may exhibit curvature singularities at the horizon \cite{Horowitz2023}. This singularity, generated by backreactions of the field, leads to a curvature singularity that grows with the size of the horizon, $r_+$, as opposed to typical tidal forces, which increase for smaller black holes. Considering backreactions with the BSW effect and longitudinal spreading would be interesting. Lastly, a string with a large $p^+$ encounters tidal forces as a shock wave, changing the string's transverse modes \cite{Dodelson2021}. The study of shocks seems relevant to this analysis.

\subsection{Warping and Compactification}

%Compact dimension, modifies string coupling $g_{eff}^2 = g^2/\text{Vol}(M_{internal})$. Mass corrections to the horizon are of order $g_{eff}$ which is considered small.

%Description of possible compactifications and the effects on amplitudes and frequency on warping
The Kerr space-time as described above can be embedded in a higher space-time to one compact dimension making loop calculations tractable-- and AdS radius, R\cite{Mousatov:2020ics}. With sufficiently large R and small string coupling leads to the supergravity theory being dual to quark-gluon plasma QCD with vanishing beta function. The warped metric is then
\begin{align}
    ds^2 = e^{2A(u)}g_{BL, \mu \nu}dx^\mu dx^\nu + e^{-2A(u)}du^2 + ds^2_{M^5}
\end{align}
For center-of-mass energy, $s$, sufficiently large, $G_N s \sim 1$, we may need higher loop corrections. In \cite{Mousatov:2020ics}, eikonal contributions are added to the interaction amplitude to calculate higher loop effects. This warping scales momenta for observers seeing non-compact momenta, so Mandelstam variable are now $s \rightarrow e^{2A(u_0)}s$ and $t \rightarrow e^{2A(u_0)}t$. Further, this also scales the string tension, 
\begin{align}
    \frac{1}{4\pi\alpha} \rightarrow \frac{e^{2A(u_0)}}{4\pi\alpha}
\end{align}
and create KK excitation with the lightest mode having a mass of $R_{AdS}e^{A(u_0)}$, where $u_0$ is the position in the compact AdS direction where the theory is localized. This warping makes the tension a free parameter and provides the extra dimension that allows for tractability the eikonal amplitude for string scattering. This warped string tension may allow for a regime near the black hole horizon where large strings are long-live. An analysis of analogous to cosmic string in a cosmological model may provide insight into quantum gravity dynamics. The strings are fairly localized in around $u = u_0$, such that $A(u_0)<<1$. This can be seen in its dressed wave function.  
\begin{align}
    \Phi = e^{ip_u\cdot x_u - \frac{1}{2}\sum_i \omega_i(e^{2A}X_i^\perp X_i^\perp - e^{-2A}U_iU_i)}
\end{align}
the frequency of strings transverse mode, $\omega_i$, is inverse proportional to the the longitudinal momentum $$\omega_n^\perp \sim \frac{n}{\alpha' p^+_U}.$$ At large $p^+_U$, the string transverse diffusion becomes large, but interestingly, and spread into the large $A(u_0) << A(u)$ region is suppressed by the larger effective frequency.

%effect of warping on drama condition
Considering the warping in the extra-dimensional throat 
\begin{align}
    ds^2 = \frac{r^2}{R^2}dx^\mu dx_\mu + \frac{R^2}{r^2}dr^2 + R^2d\Omega^2_5
\end{align}
and assuming the throat is sufficiently weakly curved
\begin{align}
    \frac{R^2}{\alpha'}>>1
\end{align}
leads to a modification of the drama condition where longitudinal spreading dominates. Assuming the position in the throat is $r = r_B > r_+\rightarrow r_B > \frac{R^2}{r_+}$
\begin{align}
    m & >\frac{r_+}{\alpha'_B} = \frac{r_+ r_B^2}{\alpha' R^2} \implies\\
    m &> \frac{R^2}{\alpha' r_+}
\end{align}
This modification relaxes the drama condition, which becomes more accessible for larger black holes.

\subsection{Comparison to Schwarzschild longitudinal spread}
%Drama conditions
For longitudinal spreading around a Schwarzschild black hole, a large amount of time separation is required to induce a significant boost. This leads to stringy interaction between late and early infallers. In an extremal Kerr near horizon background, the large boost needs only a critical and usual infaller involved. Without multiple scattering or infallers origin from infinity, critical infallers take a large amount of time to approach the centrifugal barrier at the horizon, so it seems that infallers with asymptotic origins may need a large time separation to generate a large COM energy even for EK geometry. Still, generically, with multiple scattering or origins from the horizon, like EK superradiant modes, implies drama. Just as early/late hawking modes can interact with late/early infallers, critical modes in the Kerr geometry may be sensitive to usual infallers. The connection between horizon modes and infallers through longitudinal spreading seems intrinsic to black hole physics and the understanding of microstates in a quantum gravitational context.
It seems inevitable that extremal black holes induced stringy center-of-mass energies between usual and critical trajectories. Extremal black branes endowed with superstring charges reproduce the entropy of black holes \cite{Strominger1996}\cite{Guica2009}. So, analyzing the fate of the BSW event in the D-brane description at small $gN$ or the black brane description at large $gN$ is interesting. Fundamentally, the information probes in each setup should be the same as long as the interaction survives the $gN$ scaling. What implications does this have for complimentary?  \iffalse If extremal black holes are on the effective low-energy gravity ($gN>1$) side of the gauge/gravity duality and still excite string modes (possibly through the BSW effect) in a smooth geometry, then phenomena natural to the D-brane description are present event at large t'Hooft coupling.\fi Given that D-brane interaction can probe below the string scale, the BSW effect may be an avenue to probe the quantum nature of gravity at sub-string scales.

\section{Future direction and Conclusion}
The BSW effect is relevant for string theoretic scenarios in an extremal Kerr background, making Planckian energies accessible. String spreading, especially the longitudinal direction, detunes the kinematic requirements, leading to the relevance of string theoretic interaction for the EK geometry. Effective field theory, in the context of the BSW effect, breaks down in this geometry. So, the near-horizon EK region is an interesting theoretical laboratory for studying quantum gravitation physics. 

Kerr black holes have been studied in several contexts relating to dualities, magnetic fields, cosmic string interaction, prediction of microstates, quasinormal modes, the addition of curvature terms, and more. It would be interesting to analyze if the application of the BSW to string theoretic data is consistent with other phenomena developed for spinning black holes. Lastly, since the BSW effect occurs for non-equatorial physics, generalizing the above conclusions away from the equator would be an interesting future direction.

\section{Acknowledgement}

We are grateful to Eva Silverstein, Matthew Dodelson, Byungwoo Kang, and Alex Mousatov for the extensive discussion. Also, thanks to Giuseppe Bruno De Luca for the support.

\bibliographystyle{unsrt}
\bibliography{detune}{}

\pagebreak

\section*{Appendix}
\subsection{Local Kinematics}\label{sec:appendix}
The BL coordinates obscure the scattering dynamics, so we use proper distance and three-velocities
\begin{equation}
    v^i = \frac{p^\mu e_\mu ^i}{-p^\mu e_\mu^0},
\end{equation}
to show the difference between critical trajectories, and usual trajectories near the horizon. Asymptotically infalling critical particles have only enough energy to reach the turning point with vanishing radial speed, $\dot{r} = \sqrt{E^2 - V_{\text{eff}}^2}$,  while the proper velocity is finite, $\dot{l} = \sqrt{\frac{3L_C^2}{4M^2}-\delta_C}$.  We can analyze the angle separation in terms of three velocities
\begin{align}
    v^l&=-\sqrt{1-\frac{M^2}{L_C^2}(\delta_C+\frac{L_C^2}{g_{\phi\phi}})}\\
    v^{\phi} &= \frac{M}{\sqrt{g_{\phi\phi}}}.
\end{align}
The lowest momentum critical velocity vector is \begin{equation}\label{eq:ISCOprop}(v^l, v^\phi)=(0, \frac{1}{2})\end{equation} with its turning point at ISCO. Its CPO counterpart has three-velocity \begin{equation}\label{eq:CPOprop}(v^l, v^\phi)=(\pm\frac{\sqrt{3}}{2}, \frac{1}{2}).\end{equation} Plunging usual orbits have vanishing $v^\phi$ at the horizon regardless of energy or angular momentum, so for equatorial usual orbits it is
\begin{align}
    v^l &=\pm\sqrt{1-\frac{N^2}{(E_U- \omega_H L_U)^2}(\delta_U+\frac{L_U^2}{g_{\phi\phi}})}\\
    v^\phi &= \frac{L_U N}{\sqrt{g_{\phi\phi}}(E_U-\omega_HL_U)},
\end{align}
which is $(\pm 1, 0)$ at the horizon. The longitudinal string spreading spreads in the direction of large light cone momentum, which follows the usual trajectory in this frame.

\end{document}